\documentclass[12pt]{iopart}
\usepackage{graphicx}
\usepackage{bm}
\begin{document}
\title{Entanglement sudden death and sudden birth in two uncoupled spins}
\author{E. Ferraro, A. Napoli, M. Guccione, A. Messina}
\address{Dipartimento di Scienze Fisiche ed Astronomiche, Universit\`{a}
di Palermo, via Archirafi 36, 90123 Palermo, Italy}
\begin{abstract}
We investigate the entanglement evolution of two qubits interacting
with a common environment trough an Heisenberg XX mechanism. We
reveal the possibility of realizing the phenomenon of entanglement
sudden death as well as the entanglement sudden birth acting on the
environment. Such analysis is of maximal interest at the light of
the large applications that spin systems have in quantum information
theory.
\end{abstract}
\maketitle
\section{Introduction}
The interest toward spin systems, and more in particular toward spin
dynamics in semiconductor structures, has remarkably increased in
the last few years also in connection with new emerging areas of
physics such as quantum information and computation. In this
framework it becomes a relevant subject to analyze the entanglement
behavior in spin systems in order to assess the performance of
applications for example in quantum information processing. Quite
recently it has been shown that entanglement in two qubits system
can experience sudden death and sudden birth. This phenomenon
\cite{Eberly},\cite{Yu} deserve a great attention also in
applicative contexts from quantum optical to condensed matter
systems and has been observed in laboratory in experiments with
entangled photon pairs \cite{Almeida} and atomic ensembles
\cite{Laurat}. In this paper we demonstrate the possibility of
realizing such a behavior in a system of two uncoupled spins in a
common environment.

\section{Physical system}
Let's consider a bipartite system constituted by two spins $A$ and
$B$, hereafter called central spins, that interact, with the same
coupling constants $\alpha$, with a system of $N$ uncoupled spins.
The hamiltonian model that describes such a physical situation is
\begin{equation}\label{Hamiltonian}
H=H_0+H_I
\end{equation}
with
\begin{equation}\label{interaction}
H_0=\omega(S_z+J_z),\qquad H_I=\alpha(S_+J_-+S_-J_+),
\end{equation}
where $S_z\equiv\frac{1}{2}(\sigma_z^A+\sigma_z^B)$ and
$S_\pm\equiv(\sigma_{\pm}^{(A)}+\sigma_{\pm}^{(B)})$ are spin
operators acting on the Hilbert space of the central system and
$J_z\equiv\frac{1}{2}\sum_{i=1}^N\sigma_z^i$ and
$J_\pm\equiv\sum_{i=1}^N\sigma_{\pm}^i$ are the collective operators
describing the others $N$ spins. In solid state physics, for
example, this model can effectively describe many physical systems
such as quantum dots \cite{Imamoglu}, two-dimensional electron gases
\cite{Privman} and optical lattices \cite{Sorensen}. The
time-dependent Schr\"{o}dinger equation has been already solved for
an arbitrary initial condition \cite{Palumbo}, \cite{Napoli}. In
what follows we analyze the dynamics of the entanglement in the
central system when the surrounding spins are prepared in specific
initial conditions.

\section{Collapses and revivals in the entanglement evolution}
\subsection{Binomial initial state}
Suppose that the $N$ uncoupled spins around the central system are
prepared in a linear superposition, with binomial weight, of
eigenstates $|J, M\rangle$ of $J^2$ and $J_z$ with $M=0$. The two
central spins $A$ and $B$ are instead prepared in the state
$|S=1,M_S=0\rangle$ that is a maximally entangled state. The initial
condition we are considering can be thus written as
\begin{equation}\label{2}
|\psi(0)\rangle=\sum_{J=0}^{N/2}B_0^J|1,0\rangle|J,
0\rangle\equiv\sum_{J=0}^{N/2}B_0^J|1,0,J, 0\rangle,
\end{equation}
where
\begin{equation}B_0^J=\left[{\frac{N}{2}\choose
J}p^{J}(1-p)^{\frac{N}{2}-J}\right]^{\frac{1}{2}},\qquad p\in[0,1].
\end{equation}
At a time instant $t$ we can write \cite{Palumbo}
\begin{equation}\label{psit}
\hspace{-1cm}|\psi(t)\rangle=\sum_{J=0}^{N/2}B_0^J\left(A_J(t)|1,0,J,
0\rangle-iB_J(t)|1,1, J,-1\rangle-i C_J(t)|1,-1, J, 1\rangle\right)
\end{equation}
with
\begin{equation}\label{ab}A_J(t)=\cos(2\,q_J\,\alpha t),\qquad B_J(t)=C_J(t)=\frac{1}{\sqrt{2}}\sin(2\,q_J\,\alpha t),\end{equation}
where $\label{q}q_J=\sqrt{J(J+1)}$. Exploiting
eqs.(\ref{psit})-(\ref{ab}) it is possible to prove that at any time
instant $t$ the reduced density matrix of the central system in the
two-spin standard basis
$\{|\!\!\uparrow\uparrow\rangle,|\!\!\uparrow\downarrow\rangle,|\!\!\downarrow\uparrow\rangle,|\!\!\downarrow\downarrow\rangle\}$,
has the following quite simple structure
\begin{equation}\label{struttura2}
\rho_{AB}(t)=\left(%
\begin{array}{cccc}
  b(t) & 0 & 0 & 0 \\
  0 & a(t) & a(t) & 0 \\
  0 & a(t) & a(t) & 0 \\
  0 & 0 & 0 & b(t) \\
\end{array}%
\right),
\end{equation}
where
\begin{equation}
a(t)=\frac{1}{2}\sum_{J=0}^{N/2}(B_0^{J})^2A_J(t)^2,\qquad
b(t)=\sum_{J=0}^{N/2}(B_0^{J})^2B_J(t)^2.
\end{equation}
To estimate the entanglement in the central system we adopt the
well-known concurrence function C \cite{Wootters} that in our case
can be simply expressed as
\begin{equation}\label{exact}
\hspace{-2.3cm}C(t)=\max\left[0,2\
\sum_{J=0}^{N/2}(B_0^{J})^2\left(\frac{1}{2}A_J(t)^2-B_J(t)^2\right)\right]\equiv\max\left[0,\
\sum_{J=0}^{N/2}(B_0^{J})^2\left(1-2(\Delta S_z(t))^2\right)\right],
\end{equation}
where $(\Delta S_z(t))^2\equiv\langle S_z^2\rangle(t)-(\langle
S_z\rangle(t))^2$. The presence of entanglement in the two central
spins thus is strictly related to the behavior of observable of
clear physical meaning. It is interesting to underline that for
$\alpha t\ll N$ and $p=\frac{1}{2}$, it's possible to find the
following closed form of $C(t)$
\begin{equation}\label{approximation}
C(t)\simeq\max\left[0,\cos^{\frac{N}{2}}(2\alpha t)\cos((N+2)\alpha
t)\right].
\end{equation}
\begin{figure}[h]
\begin{center}
\includegraphics[scale=0.8]{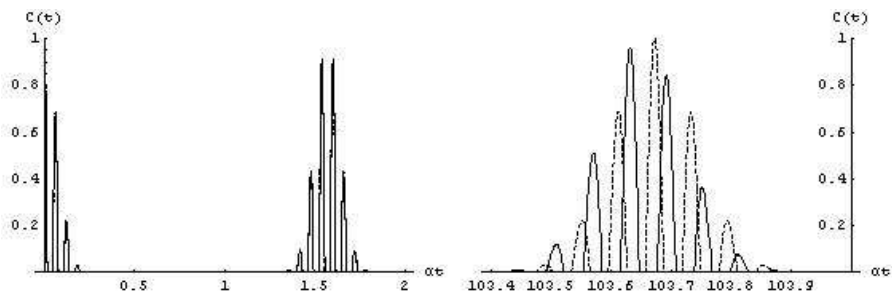}
\caption{$C(t)$ in function of $\alpha t$ for $N=100$ with $p=1/2$}
\end{center}
\end{figure}
Fig.$1$ displays the behavior of the exact concurrence function
$C(t)$ as given by eq.(\ref{exact}) (continous line) and its
approximation, eq.(\ref{approximation}), (dot line). As expected the
agreement is excellent at least for $\alpha t\ll N$. Moreover the
figure puts into light an interesting behavior in the time evolution
of the entanglement present in the central spins. Starting, indeed,
by construction from $C(0)=1$, the concurrence function evolves
showing collapses and revivals phenomena. On the other hand, it is
possible to prove that during the plateau of $C(t)$ (that is when
$C(t)$ maintains the zero value) the two spins are in a separable
state described by the following density matrix
\begin{equation}
\rho_{AB}(t)=\frac{1}{4}\left(%
\begin{array}{cccc}
  1 & 0 & 0 & 0 \\
  0 & 1 & 1 & 0 \\
  0 & 1 & 1 & 0 \\
  0 & 0 & 0 & 1 \\
\end{array}%
\right),
\end{equation}
that describes a system characterized by an equal probability of
finding all the states
$|\!\!\uparrow\uparrow\rangle,|\!\!\uparrow\downarrow\rangle,|\!\!\downarrow\uparrow\rangle,|\!\!\downarrow\downarrow\rangle$.
After the plateau of the entanglement the concurrence function
suddenly grows up reaching values near to $1$. This behavior
periodically appears. The collapses and revivals of $C(t)$ shown in
Fig.$1$ recalls those we have in the dynamical behavior of the
two-photon Jaynes-Cummings model  \cite{Joshi1},\cite{Joshi2}
described by the following Hamiltonian model
\begin{equation}\label{JC2}
H_{JC}=\frac{\hbar\omega_0}{2}\sigma_3+\hbar\omega
a^{\dagger}a+\hbar\lambda(\sigma_+a^2+\sigma_-a^{\dagger 2}),
\end{equation}
where $a$ is the annihilation operator of the single cavity mode.
The analogy between our spin star system and the two-photon J-C
model can be better brought to light following the suggestion of
ref\cite{Gerry}, that is putting
\begin{equation}\label{trasf}
J_+=\frac{a^{\dagger2}}{2},\qquad J_-=\frac{a^{2}}{2}.
\end{equation}
Exploiting indeed such a correspondence the interaction Hamiltonian
(\ref{interaction}) assumes the form
\begin{equation}
H_I=\frac{\alpha}{2}(S_+a^2+S_-a^{\dagger 2}).
\end{equation}

\subsection{Atomic coherent initial state} In this Section we analyze
a different initial condition for the $N$ spins around the central
system that is the well know \emph{atomic coherent state},
introduced in $1972$ by Arecchi in analogy with the coherent states
of the radiation \cite{Arecchi}. The central system is instead once
again in the state $|1,0\rangle$. A coherent state of $N$ spins is a
linear superposition of states $|J,M\rangle$ obtained fixing $J$ and
varying $M$. In particular, putting $J=\frac{N}{2}$, the initial
state of the global system is the following
\begin{equation}
|\psi(0)\rangle=\sum_{M=-N/2}^{N/2}B_M^{\frac{N}{2}} |\frac{N}{2},
M\rangle|1,0\rangle\equiv\sum_{M=-N/2}^{N/2}B_M^{\frac{N}{2}}|1,0,
\frac{N}{2}, M\rangle
\end{equation}
where
\begin{equation}
B_M^{\frac{N}{2}}=\left[{N\choose
M+\frac{N}{2}}p^{M+\frac{N}{2}}(1-p)^{\frac{N}{2}-M}\right]^{\frac{1}{2}},\qquad
p\in[0,1].
\end{equation}
Starting from $|\psi(0)\rangle$ at time instant $t$ we have
\cite{Palumbo}
\begin{equation}
\hspace{-3cm}|\psi(t)\rangle=\sum_{M=-N/2}^{N/2}B_M^{\frac{N}{2}}\left(A_M(t)|1,0,\frac{N}{2},
M\rangle-iB_M(t)|1,-1, \frac{N}{2}, M+1\rangle-iC_M(t)|1,1,
\frac{N}{2}, M-1\rangle\right)
\end{equation}
with
\begin{equation}\label{a}A_M(t)=\cos(\sqrt{2(q_M^2+r_M^2)}\alpha t),\qquad B_M(t)=\frac{r_M}{\sqrt{q_M^2+r_M^2}}\sin(\sqrt{2(q_M^2+r_M^2)}\alpha t),\end{equation}
\begin{equation}\label{c}C_M(t)=\frac{q_M}{\sqrt{q_M^2+r_M^2}}\sin(\sqrt{2(q_M^2+r_M^2)}\alpha t)\end{equation}
where
\begin{equation}\label{qr}q_M=\sqrt{\frac{N}{2}\left(\frac{N}{2}+1\right)-M(M-1)},\qquad r_M=\sqrt{\frac{N}{2}\left(\frac{N}{2}+1\right)-M(M+1)}.\end{equation}
In this case the concurrence function becomes
\begin{equation}
C(t)=\max\left[0,2\sum_{M=-N/2}^{N/2}(B_M^{\frac{N}{2}})^2\left(\frac{1}{2}A_M(t)^2-\sqrt{B_M(t)^2\,C_M(t)^2}\right)\right].\end{equation}
The dynamical evolution of $C(t)$ against $\alpha t$ is shown in
Fig.$2$.
\begin{figure}[h]
\begin{center}
\includegraphics[scale=0.8]{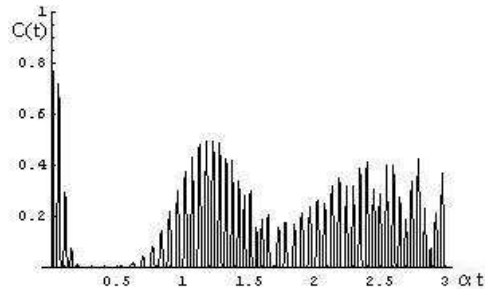}
\caption{$C(t)$ in function of $\alpha t$ for $N=100$ with $p=0.9$}
\end{center}
\end{figure}
We observe that in this case the situation is quite different from
the situation previously examined: the entanglement initially
present in the central system sudden dies after some oscillations
and, after a period of time in which it is absent, lives again.
However in this case the concurrence function does not reach values
near $1$ assuming values less than $\frac{1}{2}$. Once again it's
possible to make a parallel between the J-C model and the spin star
system exploiting the Holstein-Primakoff transformations
\cite{Primakoff}
\begin{equation}
J_+=\sqrt{2J}a^{\dagger}\sqrt{1-\frac{a^{\dagger}a}{2J}},\qquad
J_-=\sqrt{2J}\sqrt{1-\frac{a^{\dagger}a}{2J}}\,a,
\end{equation}
that are valid in a subspace with $J$ fixed. Operating such a
transformation the interaction Hamiltonian (\ref{interaction})
becomes
\begin{equation}\label{rewrite}
H_I=\alpha\sqrt{N}\left[S_+\sqrt{1-\frac{a^{\dagger}a}{N}}\,a+S_-a^{\dagger}\,\sqrt{1-\frac{a^{\dagger}a}{N}}\right],
\end{equation}
that in the limit of a large number $N$ of spins in the environment,
that is $\sqrt{1-\frac{a^{\dagger}a}{N}}\simeq 1$, reduces to an
Hamiltonian of the J-C type.

\section{Conclusion}
Summarizing in this paper we have focused our attention on a system
constituted by two uncoupled spins embedded in a common environment
composed by $N$ spins. We have proved the possibility of realizing
periodic sudden death and birth of the entanglement in the two not
interacting spins appropriately choosing the initial condition of
the environment. Generally speaking sudden death and sudden birth of
the entanglement provides an interesting resource for creation on
demand of entanglement between two qubits.

\end{document}